# LIT01-196, a metabolically stable apelin-17 analog, normalizes blood pressure in hypertensive DOCA-salt rats via a NO synthase-dependent mechanism

Short title: Metabolically stable apelin analog and hypertension

Adrien Flahault[a], Mathilde Keck[a], Pierre-Emmanuel Girault-Sotias[a], Lucie Esteoulle[b], Nadia De Mota[a], Dominique Bonnet[b], Catherine Llorens-Cortes[a*]

[a] College de France, Laboratory of Central Neuropeptides in the Regulation of Body Fluid Homeostasis and Cardiovascular Functions, Center for Interdisciplinary Research in Biology (CIRB), INSERM U1050 / CNRS UMR7241, Paris, F-75005, France

[b] Laboratoire d'Innovation Thérapeutique, UMR7200 CNRS/Université de Strasbourg, Labex MEDALIS, Faculté de Pharmacie, 74 route du Rhin, 67401 Illkirch, France.

[*] To whom correspondence should be addressed. Catherine Llorens-Cortes, Laboratory of Central Neuropeptides in the Regulation of Body Fluid Homeostasis and Cardiovascular Functions, INSERM, U1050, Collège de France, 11 place Marcelin Berthelot, 75005 Paris, France. Tel: + 33 1 44271663. Fax: + 33 1 44271691. e-mail: c.llorens-cortes@college-de-france.fr

**ABSTRACT**

Apelin is a neuro-vasoactive peptide that plays a major role in the control of cardiovascular functions and water balance, but has an *in-vivo* half-life in the minute range, limiting its therapeutic use. We previously developed LIT01-196, a systemically active metabolically stable apelin-17 analog, produced by chemical addition of a fluorocarbon chain to the N-terminal part of apelin-17. LIT01-196 behaves as a potent full agonist for the apelin receptor and has an *in vivo* half-life in the bloodstream of 28 minutes after intravenous (i.v.) and 156 minutes after subcutaneous (s.c.) administrations in conscious normotensive rats. We aimed to investigate the effects of LIT01-196 following systemic administrations on arterial blood pressure, heart rate, fluid balance and electrolytes in conscious normotensive and hypertensive deoxycorticosterone acetate (DOCA)-salt rats. Acute i.v. LIT01-196 administration, in increasing doses, dose-dependently decreases arterial blood pressure with $ED_{50}$ values of 9.8 and 3.1 nmol/kg in normotensive and hypertensive rats, respectively. This effect occurs for both via a nitric oxide-dependent mechanism. Moreover, acute s.c. LIT01-196 administration (90 nmol/kg) normalizes arterial blood pressure in conscious hypertensive DOCA-salt rats for more than 7 hours. The LIT01-196-induced blood pressure decrease remains unchanged after 4 consecutive daily s.c. administrations of 90 nmol/kg, and does not induce any alteration of plasma sodium and potassium levels and kidney function as shown by the lack of change in plasma creatinine and urea nitrogen levels. Activating the apelin receptor with LIT01-196 may constitute a novel approach for the treatment of hypertension.

**INTRODUCTION**

Hypertension is a major risk factor for cardiovascular diseases, with a prevalence of 37.2% (Brouwers et al., 2021). Optimal blood pressure (BP) control is difficult to achieve, and BP remains uncontrolled in 70% of patients treated for hypertension in the world (Walli-Attaei et al., 2020). The development of new therapeutic agents acting on new targets via different modes of action is therefore required to improve BP control.

Apelin is a (neuro)-vasoactive peptide first identified as the endogenous ligand of the human orphan APJ receptor (O'Dowd et al., 1993), a seven-transmembrane domain G-protein coupled receptor (GPCR) that has been renamed the apelin receptor (ApelinR) (O'Dowd et al., 1993; Tatemoto et al., 1998). Apelin is derived from preproapelin, a 77-aminoacid precursor, the last C-terminal 17 aminoacids (corresponding to apelin-17 or K17F) which are strictly conserved in all mammalian species studied. K17F is identified in plasma and tissues *in vivo* (Azizi et al., 2008). Apelin and its receptor are present in the brain, heart, kidney, and blood vessels (Lee et al., 2000; Reaux et al., 2001, 2002).

Apelin has an aquaretic effect, through both the central inhibition of arginine-vasopressin (AVP) release from the posterior pituitary into the bloodstream (De Mota et al., 2004), and the renal blockade of the AVP antidiuretic effect (Hus-Citharel et al., 2008, 2014; Flahault et al., 2017). Finally, apelin has vasorelaxant properties. Intravenous (i.v.) administrations of apelin decreases arterial BP via a nitric oxide (NO)-dependent mechanism in anesthetized normotensive rats (Reaux et al., 2001; Tatemoto et al., 2001; El Messari et al., 2004), and cause a NO-dependent arterial vasodilation *in vivo* in man (Japp et al., 2008). A meta-analysis (Xie et al., 2017) has reported lower plasma apelin concentrations in patients with hypertension. Given the BP-lowering effect of apelin, targeting the ApelinR may constitute a potential approach for the treatment of hypertension.

The *in vivo* half-life of K17F in the rat blood circulation is 50 seconds following i.v. administration (Flahault et al., 2021), accounting for the short-lasting (about 2 minutes) and modest (-12.8 mmHg for 3 nmol/kg) effect of i.v. K17F on BP in anesthetized normotensive rats (El Messari et al., 2004), which limits its use as therapeutic agent. We developed a metabolically stable K17F analog, LIT01-196, by the original chemical addition of a fluorocarbon chain to the N-terminal part of K17F (Gerbier et al., 2017). This chemical modification increased the *in vivo* half-life of LIT01-196 in the bloodstream to 28 minutes after i.v. and 156 minutes after subcutaneous (s.c.) administrations in conscious rats (Flahault et al., 2021) without altering its pharmacological properties in terms of the affinity for the ApelinR, inhibition of cAMP production, ApelinR internalization, ß-arrestin-2 recruitment and ERK phosphorylation (Gerbier et al., 2017). Like K17F, LIT01-196 induced a NO-dependent vasorelaxation of rat aorta pre-contracted with norepinephrine and rat glomerular arterioles pre-contracted with angiotensin II (Gerbier et al., 2017).

Together, this suggests that sustained activation of the ApelinR with LIT01-196 may constitute a potential new approach to decrease high BP in experimental models of hypertension. In agreement with this hypothesis, low apelin expression in hypertensive deoxycorticosterone acetate (DOCA) -salt rats (Zhang et al., 2006; Akcılar et al., 2013) and hypertensive patients (Xie et al., 2017) has been reported. Moreover, whether ApelinR activation still induces a NO-dependent arterial BP decrease in case of salt-dependent hypertension, remains to be determined. With this aim, we studied the effects of LIT01-196 on arterial BP, heart rate (HR), fluid balance and electrolytes, following i.v. and s.c. administrations in conscious normotensive (Sprague-Dawley, SD and Wistar-Kyoto, WKY) and hypertensive DOCA-salt rats, a salt- and volume-dependent ~~but renin-independent~~ model of hypertension, with low plasma renin levels, resistant to treatment with systemic renin-angiotensin system (RAS) blockers and exhibiting low NO-synthase activity (Lee et al., 1999; Basting and Lazartigues, 2017).

**MATERIALS AND METHODS**

**Drugs, Antibodies and Reagents.**

K17F was synthesized by GL Biochem (Shangai, China). LIT01-196 (300 mg) was synthesized by the *Laboratoire d'Innovation Thérapeutique* (CNRS UMR7200, Illkirch, France), as previously described (Gerbier et al., 2017). [$^{125}$I]-pE13F (monoiodinated on Lys$^8$ with Bolton-Hunter reagent, 2,200 Ci/mmol) was purchased from PerkinElmer (Wellesley, MA, USA). DOCA pellets (50 mg, 21-day release) were obtained from Innovative Research of America (Sarasota, FL, USA). Sodium heparinate, lithium heparinate, EDTA and N(ω)-nitro-L-arginine methyl ester (L-NAME) were obtained from SIGMA-ALDRICH (Saint-Quentin Fallavier, France). Rabbit polyclonal antibodies directed against the apelin fragment K17F were produced in the laboratory, as previously described (Reaux et al., 2001).

**Animals.**

Hypertensive DOCA-salt rats were obtained as follows: male WKY (250-300 g, 8 weeks of age) rats underwent unilateral nephrectomy under isoflurane anesthesia (Isovet ®, 3% for induction, 1-2% for maintenance anesthesia) and were implanted with a s.c. DOCA pellet (50 mg, 21-day release) in the rodent surgical facility of Charles River Laboratories (L'Arbresle, France). They were given free access to normal chow and water supplemented with 0.9% NaCl and 0.2% KCl, and were shipped to our animal facility one week later. Hypertension occurred after two weeks.

Normotensive male and female SD (250-300 g) and male WKY rats (250-300 g) were also obtained from Charles River Laboratories and were given free access to normal chow and water.

Animals were maintained under a 12 h light/12 h dark cycle. All experiments were carried out in accordance with current international guidelines for the care and use of experimental animals, and the experimental protocols were approved by the national animal ethics committee (CEEA, reference numbers 01966.02, 01962.02 and 2016-10#3672). All experiments involving animals took place in the animal facility of the Collège de France, Paris, France.

**Drug Administration, BP and HR Measurements.**

Rats were anesthetized with isoflurane (Isovet®, 3% for induction, 1-2% for maintenance), and they were pain relieved with lidocaïne (Laocaïne®, 4 mg/kg, s.c. injection). They were implanted with a right femoral artery catheter (.011"X.024"X.0065") and, for i.v. injection, with a right femoral venous catheter (.023"X.038"X.0075"), which were passed beneath the skin under isoflurane anesthesia. BP was recorded in conscious unrestrained animals after a recovery period of at least 24 hours, with a PowerLab/Labchart system (ADInstruments, Dunedin, New Zealand) connected to the arterial catheter via a pressure transducer. Mean arterial BP (MABP) and HR were calculated from the arterial pressure signal (Reaux et al., 1999). The compound was administered i.v. or s.c. (1 ml/kg in NaCl 0.9%) after baseline BP recording for at least 30 minutes. BP was measured for at least 2 hours after drug administration (longer if no return to baseline values was observed). When appropriate, BP was measured again 24 hours after administration. At the end of the experiment, animals were killed with a lethal dose of pentobarbital (Dolethal®, 150 mg/kg). Animals were randomly assigned to treatment groups. It was specified before the study that hypertensive DOCA-salt animals were excluded in case baseline MABP was lower than 150 mmHg.

**NO-Synthase Inhibition.**

To determine the effect of NO synthase inhibition on LIT01-196 hypotensive effect, we recorded BP following i.v. administration of L-NAME, a NO synthase inhibitor, at the dose of 30 mg/kg (111 µmol/kg), either together with or 10 minutes after the administration of LIT01-196, both in normotensive and hypertensive DOCA-salt rats, after a stabilization period of at least 30 minutes.

**Angiotensin converting enzyme 2 (ACE2) enzymatic activity measurement**

ACE2 enzymatic activity was determined on cardiac homogenates and the fluorogenic substrate (7-methoxycoumarin-4-yl)acetyl-Mca-Ala-Pro-Lys(2,4-dinitrophenyl)(Dnp)-OH (Mca-APK(Dnp), as previously described (Pedersen et al., 2011; Ye et al., 2012). The rats received saline or LIT01-196 (90 nmoles/kg, s.c., for 4 days), under a volume of 250µl. We used five rats for each set of conditions. After sacrifice of the animals by decapitation, hearts were removed without the atria, rinsed in cold saline, frozen in liquid nitrogen and then stored at -80 °C. The hearts were then thawed in 6ml of cold ACE2 buffer (50 mM 4-morpholineethanesulfonic acid, 300 mM NaCl, 10 µM ZnCl2, and 0.01% Triton-X-100, pH 6.5) and homogenized by sonication. Aliquots of the heart homogenates (20 µl) were added to wells containing 10 μM Mca-APK(Dnp) (10μl), in the presence or absence of 1 μM MLN4760, a specific and selective ACE2 inhibitor (10 μl) in a total volume of 100 μl ACE2 buffer. Plates were incubated for 30 min at 37°C and subjected to repeated reading with the Victor Nivo multilabel reader (PerkinElmer), with fluorescence emission at 405 nm and excitation at 320 nm. To estimate the specific ACE2 activity, the arbitrary fluorescence units from the assay with MLN4760 was subtracted from the fluorescence units without the inhibitor. The activities were expressed in RFU per μg protein per min."

**Apelin Radioimmunoassay (RIA).**

Trunk blood was collected after decapitation of the animal on ice (50 μl of 0.3 M EDTA (pH 7.4) for 1 mL of blood). Plasma samples (0.5 ml) were acidified with 0.175 ml of 3 M HCl and stored at -80°C until apelin RIA. After thawing the sample on ice, 0.05% BSA was added and the samples were centrifuged at 20,000 x *g* and 4°C for 10 min. The supernatants were collected and the pH was adjusted to 6.5 with 10 M NaOH and 2 M Tris-HCl buffer (pH 7.4). Apelin was extracted from plasma by mixing 0.3 ml of the supernatant with 0.3 ml of 1% trifluoroacetic acid (TFA) – 0.1% BSA, and loading it onto a Sep-Pak C18 cartridge (Waters, Massachusetts, USA) previously washed with 2 ml 100% acetonitrile and equilibrated with 5 ml 1% TFA – 0.1% BSA. The columns then were washed with 3 ml 1% TFA - 0.1% BSA and apelin was eluted with 1.5 ml 100% acetonitrile. The samples were dried and dissolved in 0.32 ml of RIA buffer (19 mM $NaH_2PO_4 \cdot H_2O$, 81 mM $Na_2HPO_4 \cdot 2H_2O$, 50 mMNaCl, 0.1% TritonX-100, 0.01% $NaN_3$, 0.1% BSA).

Plasma apelin levels were determined by RIA on 0.1 ml of plasma, with a polyclonal K17F antiserum (0.05 ml at a dilution of 1:4,500) and $^{125}$I-labeled pE13F (iodinated on Lys$^8$ by the Bolton and Hunter method, 2,200 Ci/mmol, Perkin-Elmer, Waltham, MA; 0.05 ml, 19,000 dpm) as a tracer, incubated at 4°C overnight. Samples were mixed with $^{125}$I-labeled pE13F and polyclonal K17F antiserum (dilution: 1:4,500) to give a total volume of 0.2 ml, and were incubated at 4°C overnight. We then added 0.5 ml of Amerlex (Amersham RPN 510), and the resulting mixture was incubated for 10 min at room temperature. The tubes were centrifuged at 2,600 x *g* at 4°C for 20 min. The supernatant was removed and the radioactivity of the precipitates was measured. We assessed the cross-reactivity of the apelin antiserum with various N- and C-terminally truncated fragments of K17F and several other bioactive peptides. K17F (200% cross-reactivity), the pyroglutamyl form of apelin-13 (pE13F) and apelin-36 (100% cross-reactivity) were well recognized by the antiserum, whereas the removal of the phenylalanine residue at the extreme end of the C-terminus of K17F (forming K16P) decreased recognition to barely detectable levels (<0.3% cross-reactivity). Negligible cross-reactivity was observed for angiotensin II, angiotensin III, neuropeptide Y and AVP. This antiserum identified the apelin present in the plasma of rodents as pE13F and, to a lesser extent, K17F (De Mota et al., 2004). Most of the apelin in human plasma was K17F, followed by pE13F, and, to a lesser extent, apelin-36 (Azizi et al., 2008).

**Metabolic Studies.**

For metabolic studies, animals were individually housed in metabolic cages (Techniplast). After acclimatization over a period of 24 hours, animals received an i.v. or s.c. injection of the compound or of an equivalent volume (1 ml/kg) of normotonic saline (NaCl 0.9%). They were kept in metabolic cages for 24 hours, during which time water intake and diuresis were measured. Twenty-four hours after drug administration, urine samples were collected for the determination of urine osmolality and electrolytes ($Na^+$, $K^+$, $Cl^-$), blood was collected by intracardiac puncture (10 µl of lithium heparinate (1600 IU/ml) per ml of blood) under isoflurane anesthesia, and the animals were immediately killed by carbon monoxide inhalation. Plasma and urine electrolytes were determined with an ISE 3000 analyzer (Caretium Medical Instruments, Shenzhen, China). Urine osmolality was determined with a Cryobasic 1 osmometer (AstoriTecnica, Poncarale, Italy). Plasma and urine urea, creatinine and glucose were determined using a Konelab 20I analyzer (ThermoFisher Scientific, Illkirch, France). Water excretion fraction was determined as (Plasma creatinine / Urine creatinine) * 100.

**Statistical Analysis.**

Comparisons between groups were performed with *t*-tests (2 groups) or ANOVA followed by Dunnett's or Sidak's test (>2 groups). For the comparison of repeated measurements over time, we used a linear mixed effects model with time and the measured parameter as fixed effects and the unique animal identification number as a random effect, followed by Dunn's post-hoc comparison, using the baseline value of each group as the reference. Data are represented as the means, with error bars to indicate the standard error of the mean (SEM). Each group included at least 4 separate measurements. Statistical analysis was conducted with R software version 3.4.3 (Team R Core, 2018) and additional packages (Wickham, 2009; Ritz et al., 2015).

**RESULTS**

**LIT01-196 Intravenous Administration Decreases BP and Increases HR in Conscious Normotensive SD and WKY Rats**

In conscious normotensive SD rats, basal MABP was 112 ± 1 millimeters of mercury (mmHg) and basal HR was 405 ± 8 beats per minute (bpm). The i.v. administration of a supra maximal dose of 400 nmol/kg K17F in conscious SD rats led to a maximal decrease in MABP of 12 ± 4 mmHg, lasting approximately 10 minutes (**Figure 1A**). In contrast, the i.v. administration of LIT01-196, in conscious SD rats at a much lower dose (15 nmol/kg), was associated with a maximal decrease in MABP of 59 ± 3 mmHg at 10 minutes. This hypotensive effect lasted for up to 45 minutes (**Figure 1A**). K17F (400 nmol/kg) and LIT01-196 (15 nmol/kg) induced a compensatory increase in HR of 111 ± 33 and 100 ± 19 bpm, respectively, and the effect on HR lasted longer for LIT01-196 (65 min) than for K17F (20 min) (**Figure 1B**). The i.v. administration of LIT01-196 in increasing doses in conscious SD male and female rats elicited a dose-dependent decrease in MAPB, with $ED_{50}$ (median effective dose) values of 9.8 ± 0.5 nmol/kg and 6.9 ± 0.5 nmol/kg, respectively (**Figure 1C and Supplemental Figure 1**). The i.v. administration of LIT01-196 in increasing doses in conscious normotensive male WKY rats, also induced a dose-dependent decrease in MABP (**Figure 1D**) with a significantly 1.7 fold lower $ED_{50}$ value (5.7 ± 0.4 nmol/kg, **Figure 1C**) than the value obtained for normotensive SD male rats of the same age and different genetic backgrounds. LIT01-196 (14 nmol/kg) also induced an increase in HR in male WKY rats (maximal increase of 74 ± 12 bpm, **Figure 1E**)

**LIT01-196 Intravenous Administration Decreases BP but does not Change HR in Conscious Hypertensive DOCA-Salt Rats**

We then evaluated the hemodynamic effects of LIT01-196 in conscious male hypertensive DOCA-salt rats derived from the WKY strain. In hypertensive rats, baseline MABP was 176 ± 3 mmHg and baseline HR was 403 ± 7 bpm. The i.v. administration of LIT01-196 at 15 nmol/kg led to a maximal MABP decrease of 125 ± 1 mmHg, 12 minutes after LIT01-196 administration (**Figure 1F**). LIT01-196 acute treatment (15 nmol/kg, i.v.) did not significantly modify HR (maximal increase of 11 ± 14 bpm) (**Figure 1G**). At this dose, MABP was normalized (around 100 mmHg), remained stable between 2 and 6 hours after i.v. administration and returned to baseline value after 24 hours (157 ± 8 vs 164 ± 6 mmHg, $p$=0.5) (**Figure 1H**), whereas HR was not modified (**Figure 1I**). Lower doses of LIT01-196 resulted in a smaller initial hypotensive effect, with a shorter effect on MABP (**Figure 1F**). The i.v. injection of 3 nmol/kg LIT01-196 into hypertensive DOCA-salt rats induced a decrease in MABP (maximal decrease of 60 ± 10 mmHg, **Figure 1F**) with no change in HR (maximal increase of 4 ± 7 bpm, **Figure 1G**), whereas the i.v. administration of LIT01-196 at 4 nmol/kg into normotensive WKY rats did not decrease MABP (maximal decrease of 4 ± 3 mmHg, **Figure 1D**). LIT01-196, i.v. administered in increasing doses, in conscious hypertensive DOCA-salt rats dose-dependently decreased MABP, with a significantly lower $ED_{50}$ value than that obtained for normotensive WKY rats of the same age and genetic background ($ED_{50}$ value: 3.1 ± 0.6 nmol/kg versus 5.7 ± 0.4 nmol/kg, $p$=0.007) (**Figure 1C**). In contrast, the i.v. administration of K17F at a dose of 20 nmol/kg in conscious hypertensive DOCA-salt rats did not significantly modify MABP and HR (**Figure 2A and 2B**).

**BP Decrease Induced by LIT01-196 Intravenous Administration is Inhibited in the Presence of a NO-Synthase Inhibitor in Normotensive and Hypertensive DOCA-Salt Rats.**

When LIT01-196 (15 nmol/kg) and L-NAME (111 µmol/kg) were concomitantly administered *via* the i.v. route in conscious normotensive SD rats, the decrease in BP induced by LIT01-196 was drastically reduced. Only a transient initial decrease in MABP (maximal decrease in MABP of -19 ± 4 mmHg) lasting for less than 1 minute was observed (**Figure 3A**). We found similar results in hypertensive DOCA-salt rats, in which L-NAME fully inhibited the BP decrease induced by LIT01-196 (3 nmol/kg) when the two compounds were concomitantly administered by i.v. route (**Figure 3B**).

**Prolonged and Long-Lasting BP Decrease Induced by LIT01-196 Subcutaneous Administration in Conscious Normotensive WKY Rats.**

With a view to prevent the intense initial decrease in BP induced by i.v. LIT01-196 administration and to increase the duration of action of this peptide, we first investigated the effects of LIT01-196 s.c. administration on MABP and HR in normotensive WKY rats. The s.c. administration of 90 nmol/kg LIT01-196 in conscious normotensive WKY rats resulted in a maximal decrease in MABP of 30 ± 8 mmHg (**Figure 4A**), and did not alter HR (**Figure 4B**). The BP decrease was maximal at 120 minutes and persisted at least 6 hours after the s.c. administration (**Figure 4A**). The dose of LIT01-196 (90 nmol/kg) given by s.c. route inducing a maximal BP decrease of 30 ± 8 mmHg was 11-fold higher than that required (8 nmol/kg) to have a similar hypotensive effect after i.v. LIT01-196. The duration of the hypotensive effect induced by LIT01-196 (90 nmol/kg) after s.c. administration was significantly longer (at least 6-fold) than that recorded after i.v. administration (8 nmol/kg) (**Figure 1D and Figure 4A**).

**Prolonged and Long-Lasting Antihypertensive Action of LIT01-196 after Acute or Chronic Subcutaneous Administration in Conscious Hypertensive DOCA-Salt Rats.**

We then evaluated the effects of LIT01-196 administered via the s.c. route in conscious hypertensive DOCA-salt rats. Following the s.c. injection of 90 nmol/kg LIT01-196, a decrease in MABP was observed that was smoother than after i.v. injection (**Figure 4C**). At the time of maximal effect, 20 minutes after s.c. administration, MABP was 97 ± 14 mmHg (**Figure 4C**). The s.c. administration of 90 nmol/kg LIT01-196 allowed a prolonged normalization of MABP, which remained between 97 ± 14 and 136 ± 15 mmHg from the injection until the end of recording, seven hours later (**Figure 4C**). MABP returned to baseline levels, 24 hours after the s.c. injection of 90 nmol/kg LIT01-196 (174 ± 14 vs. 187 ± 9 mmHg, *p*=0.84). The dose of LIT01-196 (90 nmol/kg) given by s.c. route, normalizing BP around 100 mmHg (**Figure 4C**) was 6 fold higher than that of LIT01-196 (15 nmol/kg) given by i.v. route in DOCA-salt hypertensive rats (**Figure 1H**). Subcutaneous administration of 90 nmol/kg LIT01-196 did not significantly increase HR, and was followed by a significant decrease in HR from the 2$^{nd}$ hour to the 4$^{th}$ hour following administration (maximal decrease of 99 ± 18 bpm) (**Figure 4D**).

The antihypertensive effect of LIT01-196 remained after 4 consecutive daily s.c. injections of 90 nmol/kg of the compound with a BP decrease that lasted at least 6 hours, and a maximal decrease in MABP of 59 ± 8 mmHg (**Figure 5A**), without modification of HR (**Figure 5B**).

**Cardiac ACE2 activity after Chronic Subcutaneous Administration of LIT01-196 in Conscious Hypertensive DOCA-Salt Rats**

Cardiac ACE2 activity in hypertensive DOCA-salt rats is decreased by 37% compared to cardiac ACE2 activity measured in control WKY rats (10.3 ± 0.8 RFU / min / µg protein, n=5 vs 16.4 ± 0.6 RFU / min / µg protein, n=5, p<0.0001). LIT01-196 given during four consecutive days at the dose of 90 nmol/kg by s.c. route to DOCA-salt rats does not modified cardiac ACE2 activity compared to that measured in DOCA-salt rats treated with saline (10.8 ± 0.3 RFU / min / µg protein, n=5 vs 10.3 ± 0.8 RFU / min / µg protein, n=5). Cardiac ACE2 activity in DOCA-salt rats treated with LIT01-196 was significantly lower than that measured in the control WKY group (10.3 ± 0.8 RFU / min / µg protein, n=5 vs 16.4 ± 0.6 RFU / min / µg protein, n=5).

**Plasma Apelin Levels in Normotensive and Hypertensive DOCA-Salt Rats**

Plasma apelin levels were lower in hypertensive DOCA-salt rats than in normotensive WKY rats of the same age and genetic background (0.509 ± 0.04 versus 0.781 ± 0.10 pmol/ml respectively, *p*=0.029) (**Figure 6**).

**Metabolic Effects of Repeated LIT01-196 Subcutaneous Administrations in Hypertensive DOCA-Salt Rats**

We then evaluated the effects of repeated doses of LIT01-196 (daily s.c. administrations of 90 nmol/kg for 4 days) on metabolic parameters in DOCA-salt rats housed in metabolic cages. After chronic treatment for 4 days, LIT01-196 did not induce any significant change in urine output, urine osmolality, water and food intake, plasma sodium and potassium levels, kidney function (plasma creatinine and urea nitrogen), plasma glucose levels and water excretion fraction versus DOCA-salt rats receiving saline (**Figure 7A-J**). Animals treated with LIT01-196 had a significant 31% higher daily urinary sodium excretion (*p*=0.047) and a similar daily urinary potassium excretion when compared to DOCA-salt rats receiving saline (**Figure 7K-L**).

**DISCUSSION**

This study describes a way of normalizing BP in an experimental salt-dependent model of hypertension, by activating the ApelinR, using a systemically active metabolically stable K17F analog, LIT01-196. We first show the hypotensive effect of an acute LIT01-196 administration by i.v. and by s.c. route in normotensive rats. We then show that acute LIT01-196 administration by s.c. route at a dose in the nmol/kg range normalizes arterial BP in conscious hypertensive DOCA-salt rats for at least 7 hours, via a NO-dependent mechanism. The antihypertensive effect of LIT01-196 is maintained after 4 consecutive days of treatment, and does not alter plasma sodium and potassium levels and kidney function as shown by the lack of change in plasma creatinine and urea nitrogen levels. This shows for the first time that LIT01-196, a metabolically stable analog of the endogenous peptide K17F acts as a potent and long-lasting antihypertensive agent.

We first found that, LIT01-196 administered in increasing doses by the i.v. route in conscious normotensive male and female SD and male WKY rats decreased BP in a dose-dependent manner, with $ED_{50}$ values of 9.8, 6.9 and 5.7 nmol/kg respectively, showing a slight higher sensitivity of female SD and male WKY rats to LIT01-196 treatment compared to male SD rats. These data on male and female SD rats are to put in line with data showing in healthy volunteers higher age-adjusted plasma apelin concentrations in men than in women (Blanchard et al., 2013). We have then pursued our animal studies with male normotensive and hypertensive DOCA-salt rats. However, it would be important to further investigate the effects of LIT01-196 on BP in female as compared to male hypertensive DOCA-salt rats since a male susceptibility and female resilience to developing hypertension have been observed in the vast majority of animal models of hypertension, regardless of whether it is induced or genetic (Arnold et al., 2017).

The i.v. administration of LIT01-196 (14-15 nmol/kg) in conscious male WKY and SD rats elicited a rapid, marked decrease in BP (-75 and -59 mmHg respectively) that was sustained for at least 60 and 45 minutes respectively whereas, at the same dose, K17F induced no significant change in BP. Nevertheless, when administered i.v. at a dose of 400 nmol/kg (27 times higher) in conscious SD rats, K17F induced a -12 mmHg decrease in BP that lasted 10 minutes. These results are in line with the previously reported *in vivo* LIT01-196 half-life of 28 minutes in the bloodstream, versus 50 seconds for K17F, in SD male rats after i.v. administration (Flahault et al., 2021). The LIT01-196-induced BP decrease does not result from the action of the compound at the brain level since LIT01-196 was previously shown not to cross the blood-brain barrier and enter the brain after systemic administration to control rats (Flahault et al., 2021). Other metabolically stable apelin analogs have been developed to increase the half-life of the peptide and its *in vivo* activity. *In vivo* effects on BP have been evaluated for only a few of these analogs. Murza *et al.* showed that a Tyr(OBn) apelin-13 analog, administered i.v. at a dose of 620 nmol/kg in anesthetized normotensive rats, induced a higher BP decrease (-50 mmHg, return to baseline after 45 minutes) compared to pE13F (Murza et al., 2015). Wang *et al.* designed the Nle-Aib-BrF-pyr-apelin-13 peptide, which decreased BP by -25 mmHg for more than 120 minutes (no return to baseline MABP was observed) when administered i.v. at a high dose (1400 nmol/kg) in normotensive anesthetized mice (Wang et al., 2016). Thus, at a dose lower than those used in these previous studies (by a factor of 41 to 93), and in conscious rather than anesthetized rats, LIT01-196 was more active than the apelin-13 analogs. LIT01-196 may, therefore, be considered as a potent hypotensive apelin analog *in vivo*.

We then investigated the effects of LIT01-196, on BP in conscious hypertensive DOCA-salt rats, a salt- and volume-dependent ~~but renin-independent (low plasma renin levels)~~ model of hypertension, with low plasma renin levels, resistant to treatment with systemic RAS blockers (Morton et al., 1979). The DOCA-salt rat has a high BP, with high plasma AVP levels, water and salt retention, hyperactivity of the brain RAS but a low activity of the systemic RAS (Basso et al., 1981; Ganten et al., 1983). We first showed that plasma apelin levels were lower in DOCA-salt rats than in normotensive rats, consistent with low apelin expression in experimental models of hypertension (Zhang et al., 2006; Akcılar et al., 2013) and hypertensive patients (Xie et al., 2017). We then found that LIT01-196 i.v. administered in increasing doses to conscious DOCA-salt rats, potently decreased BP in a dose-dependent manner, with an $ED_{50}$ value of $3.1 \pm 0.6$ nmol/kg, significantly lower than the value obtained for normotensive WKY rats of the same age and genetic background. The i.v. administration of LIT01-196 (15 nmol/kg) in conscious DOCA-salt rats elicited a high decrease in BP (-125 mmHg), versus - 75mmHg in normotensive WKY rats, that was sustained for 6 hours versus 1 hour in WKY rats. Hypertensive DOCA-salt rats appear to be more sensitive to the antihypertensive effects of LIT01-196 treatment than normotensive WKY rats. Indeed, following the i.v. administration of LIT01-196 at a dose of 3 nmol/kg, MABP decreased markedly (-60 mmHg) for at least 6 hours in DOCA-salt rats but remained unchanged in normotensive WKY rats. Similarly, Lee et al. reported that, for the same dose of apelin-13 (10 nmol/kg), the decrease in MABP was stronger in anesthetized hypertensive SHRs (-60%) than in normotensive WKY rats (-30%) (Lee et al., 2005). They also reported a short duration of the effects of apelin (6 min) in SHR.

ACE2 was previously shown to be involved in the metabolism of K17F (Wang et al., 2016) and LIT01-196 (data not shown). Since a decrease in renal ACE2 mRNA expression and brain ACE2 enzymatic activity have been observed in hypertensive DOCA-salt rats compared to normotensive rats (Crackower et al., 2002; Hmazzou et al., 2021), we have shown, in the present study, that cardiac ACE2 activity in hypertensive DOCA-salt rats was decreased compared to that measured in normotensive WKY rats. Moreover, LIT01-196 given by s.c. route during four days at the dose of 90 nmol/kg did not modified cardiac ACE2 activity compared to that measured in DOCA-salt rats treated with saline but was significantly lower than that measured in control normotensive rats. This shows that metabolism of LIT01-196 by ACE2 was decreased in hypertensive rats compared to normotensive rats and could be responsible for higher LIT01-196 levels available for ApelinR activation. This could explain the higher sensitivity of hypertensive DOCA-salt rats to LIT01-196 treatment than normotensive rats. Moreover, Akcılar *et al.* (Akcılar et al., 2013) showed in DOCA-salt hypertensive rats, that plasma apelin levels and apelin-APJ mRNA expression levels were reduced. Consequently treatment with LIT01-196 may compensate the decrease activity of the vascular apelinergic system and could participate for normalization of arterial BP. Considering the low activity of the systemic RAS in DOCA-salt rats (Lee et al., 1999; Basting and Lazartigues, 2017) and the fact that apelin inhibits angiotensin II-induced BP increase (Ishida et al., 2004), the dose of LIT01-196 required to counterbalance the stimulatory effect of angiotensin II on BP in hypertensive DOCA-salt rats is weaker than that required in normotensive rats. Finally, an impairment of the baroreflex has been reported in DOCA-salt rats (Takeda et al., 1988). In this study, despite the large amplitude of the BP decrease induced by LIT01-196 in DOCA-salt rats, HR was not increased. This could account for the higher sensitivity of DOCA-salt rats to the antihypertensive effect of LIT01-196, due to a lack of counteracting resistance adjustment by the baroreflex.

With a view to preventing the intense initial decrease in BP induced by LIT01-196 and to increase the duration of action of this peptide, we investigated the hypotensive response to the s.c. administration of this compound in normotensive and hypertensive rats. As expected, the dose required to induce a decrease in BP in conscious normotensive and hypertensive rats was larger for administration via the s.c. route that for the i.v. route. In normotensive WKY rats, the s.c. administration of 90 nmol/kg LIT01-196 induced a decrease in BP (- 30 mmHg) that was smaller than that induced by its i.v. administration, but this effect lasted for at least 6 hours, as opposed to 45 minutes for the effects of i.v.administration in normotensive animals. These results are in line with the prolonged *in vivo* half-life of LIT01-196 (156 min) in the bloodstream after s.c. administration in normotensive rats as previously reported (Flahault et al., 2021).

In cases of chronic, severe or resistant hypertension, the objective is to obtain a long-lasting effect with a progressive decrease in BP. The acute s.c. administration of LIT01-196 at a dose of 90 nmol/kg in conscious hypertensive DOCA-salt rats with a basal MABP around 176 mmHg, normalized MABP at about 100-120 mmHg for at least 7 hours without the initial drastic decrease in BP observed after i.v. administration. Despite this sustained decrease in BP, MABP measured 24 hours after the s.c. administration of LIT01-196 s.c. had returned to levels similar to those at baseline before treatment. In addition, repeated daily s.c. administrations of LIT01-196 did not alter the antihypertensive effect of the compound, and were not associated with alterations in plasma sodium and potassium levels and in kidney function. This suggests that there is no tolerance to the antihypertensive effect of LIT01-196 and no desensitization of ApelinR after chronic LIT01-196 administration. This is in agreement with the results of Akcılar *et al*. who showed that repeated intraperitoneal injection of apelin-13 for 17 days in DOCA-salt hypertensive rats, did not modified ApelinR mRNA expression in vascular tissue, relative to DOCA-salt rats receiving saline (Akcılar et al., 2013). Overall, these results indicate that LIT01-196, a metabolically stable analog of the endogenous peptide K17F, acts as a potent and long-lasting antihypertensive agent.

We also investigated the mechanism of action of LIT01-196 in the DOCA-salt model. K17F (Gerbier et al., 2017) and pE13F (Tatemoto et al., 2001; Japp et al., 2008) decrease BP through a NO-dependent mechanism in normotensive conditions, both in rodents and humans. In line with these results, we demonstrated *in vivo* that the BP decrease induced by LIT01-196 is fully inhibited by a NO synthase inhibitor, L-NAME in conscious normotensive rats. This is in agreement with the vasorelaxant activity of LIT01-196 on isolated rat aorta that were precontracted by incubation with noradrenaline and on glomerular arterioles that were precontracted with angiotensin II. In both preparations, LIT01-196, as K17F, induced a dose-dependent relaxation of the vessels (Gerbier et al., 2017), consistent with the presence of ApelinR binding sites in human aorta (Katugampola et al., 2002) and of ApelinR mRNA in the endothelial cells lining the large conductance vessels of various organs (Hus-Citharel et al., 2008; Japp and Newby, 2008). The relaxant effects of K17F and LIT01-196 on rat aorta were significantly inhibited by prior treatment of the vessels with L-NAME, suggesting that LIT01-196, similar to K17F, acted via a NO-dependent mechanism (Gerbier et al., 2017).

However, BP regulation by NO may be altered in hyperaldosteronism conditions. Studies have previously shown in DOCA-salt and 2-kidneys-1-clip rats, which are respectively models of primary and secondary hyperaldosteronism (Hashimoto et al., 1983), that plasma nitrite/nitrate ratio and aortic expression of endothelial NO synthase were decreased (Lee et al., 1999). In contrast, SHR rats exhibit an increase in NO synthase activity (Lee et al., 1999). It was thus critical to assess whether activating ApelinR by LIT01-196 in hypertensive DOCA-salt rats was also efficient to decrease BP through a NO-dependent mechanism. We show that, similarly to normotensive rats, the antihypertensive effect of LIT01-196 is fully inhibited by L-NAME in DOCA-salt rats, suggesting that LIT01-196, by activating ApelinR, decreases BP in DOCA-salt rats *via* a NO-dependent mechanism.

The fast-acting nature of the hypotensive effect of LIT01-196 suggests it mainly occurs through vasodilation. However, whether extracellular volume reduction could contribute to a longer-acting effect remained to be investigated. Indeed, the i.v. injection of K17F has been shown to increase aqueous diuresis in anesthetized lactating female rats (Hus-Citharel et al., 2008), and LIT01-196 given by i.v. or s.c. route in control rats increases urine output and decreases urine osmolality (Flahault et al., 2021). We therefore evaluated the effects of LIT01-196 administration on diuresis and urinary sodium excretion in hypertensive DOCA-salt rats after 4 days of treatment. The repeated s.c. administration of 90 nmol/kg LIT01-196 in conscious DOCA-salt rats induced a slight increase in urinary sodium excretion, with no significant increase in urine output. In contrast with prior studies in normotensive rats (De Mota et al., 2004; Hus-Citharel et al., 2014; Flahault et al., 2021), we did not observe an aquaretic effect of the compound, indeed, urine osmolality was similar in DOCA-salt rats treated with saline and LIT01-196. Due to the fact that LIT01-196 does not enter the brain and cannot inhibit AVP release in the bloodstream, the high circulating AVP levels (Morris et al., 1981; Keck et al., 2019), which promote water reabsorption by the kidney, may in turn inhibit the ability of LIT01-196 to induce water excretion.

**CONCLUSION**

In conclusion, LIT01-196, a metabolically stable analog of the endogenous peptide K17F, is a potent and long-lasting antihypertensive agent that can be administered intravenously to decrease BP rapidly, in a dose-dependent manner, or subcutaneously, to obtain a sustained and long-lasting effect. This agent acts as a vasodilator, via a NO-dependent mechanism, and, although associated with an aquaretic effect in normotensive but not in hypertensive DOCA-salt rats, it has no impact on plasma sodium or potassium levels and kidney function in DOCA-salt rats. Treatment by LIT01-196 could constitute a new approach to regulate high BP via activation of the ApelinR.

## ABBREVIATIONS

ACE2, angiotensin converting enzyme 2; ApelinR, apelin receptor; AVP, arginine-vasopressin; BP, blood pressure; DOCA, deoxycorticosterone acetate; HR, heart rate; K17F, apelin-17; L-NAME, N(ω)-nitro-L-arginine methyl ester; MABP, mean arterial blood pressure; NO, nitric oxide; pE13F, pyroglutamyl form of apelin-13; RAS, renin-angiotensin system; SD, Sprague-Dawley; WKY, Wistar-Kyoto.

## ACKNOWLEDGMENTS

Special thanks goes to the animal facility of *Collège de France* and its staff for their help and guidance. The English text was edited by J. Sappa from Alex Edelman & Associates.

## DATA AVAILABILITY STATEMENT

All supporting data are included within the main article. Raw data is available upon request to the corresponding author.

## AUTHOR CONTRIBUTION

CLC and AF contributed to the conception and design of the studies. Experimental work was carried out by AF, MK, PEGS and NDM. Data analysis was performed by AF, MK, PEGS. LIT01-196 design was performed by DB and LIT01-196 synthesis was performed by LE. CLC, AF and MK wrote the first draft of the manuscript. Resources and funding acquisition CLC and DB. All authors read and approved the submitted version.

## FUNDING

This work was supported by the French National Institute for Health and Medical Research (INSERM) [Annual dotation], including the financial support for Proof of Concept, CoPoc from INSERM Transfert, the CNRS, the *Université de Strasbourg*, the *Collège de France*, the *Agence Nationale de la Recherche* "Vie, santé et bien-être 2016" (ANR-16-CE18-0030, FluoroPEP) and the *Federation Française de Cardiologie*. The LabEx MEDALIS (ANR-10-LABX-0034) was used for the purchase of HPLC equipment. LE was supported by a fellowship from the *Ministère de l'Education Nationale, de l'Enseignement Supérieur et de la Recherche*. AF was supported by the fellowship from INSERM (Poste d'Accueil pour

Hospitaliers). PEGS was supported by a fellowship from the *Fondation pour la Recherche Médicale*, grant number "PBR201810007643".

**Conflict of Interest:** The authors declare that there are no competing interests associated with the manuscript.

**Contribution to the Field Statement**

Hypertension is a major risk factor for cardiovascular diseases, with a prevalence of 35%. Optimal blood pressure (BP) control is difficult to achieve, and BP remains uncontrolled in 70% of patients treated for hypertension in the world. The development of new therapeutic agents acting on new targets via different modes of action is therefore required to improve BP control. Apelin is a endogenous vasoactive peptide which decreases BP but its *in vivo* half-life in the minute range limits its therapeutic use. LIT01-196, a metabollically stable apelin-17 analog with an *in vivo* half-life of 156 minutes after subcutaneous administration in normotensive rats, displays a high affinity for the ApelinR and activates this receptor similarly to apelin-17. In this study, we showed that acute subcutaneous LIT01-196 administration in hypertensive DOCA-salt rats normalizes BP during more than 7 hours through the production of nitric oxide. The LIT01-196-induced BP decrease remains unchanged after 4 consecutive daily administrations, and does not induce any alteration of plasma sodium and potassium levels and kidney function. Overall, this study supports that activation of the ApelinR with a metabollically stable apelin-17 analog such as LIT01-196 but with improved pharmacokinetics could constitute a potential therapeutic approach in hypertension.

*Flahault et al. Stable apelin analog and hypertension*

**Figure legends**

**FIGURE 1. Hemodynamic effects of the intravenous administration of apelin-17 (K17F) or LIT01-196 in conscious normotensive and hypertensive DOCA-salt rats.** Effects of the i.v. administration of saline, K17F (400 nmol/kg) or LIT01-196 (3, 10.5, and 15 nmol/kg) in conscious normotensive SD rats on **(A)** MABP and **(B)** HR recorded for 1 hour and respectively expressed in millimeters of mercury (mmHg) and beats per minute (bpm). **(C)** Dose-response curve of LIT01-196 administered by the i.v. route, for MAPB changes in conscious normotensive SD, WKY and hypertensive DOCA-salt rats. Effects of the i.v. administration of saline, or LIT01-196 (4, 8, and 14 nmol/kg) in conscious normotensive WKY rats on **(D)** MABP and **(E)** HR recorded for 1 hour and respectively expressed in mmHg and bpm. **(F)** MABP and **(G)** HR recorded for 1 hour after i.v. administration of saline into conscious normotensive WKY and hypertensive DOCA-salt rats or LIT01-196 (0.6, 3, and 15 nmol/kg) into conscious hypertensive DOCA-salt rats. **(H)** MABP and **(I)** HR recorded for 6 hours and 24 hours after i.v. administration of saline into conscious normotensive WKY and hypertensive DOCA-salt rats or LIT01-196 (15 nmol/kg) into conscious hypertensive DOCA-salt rats. Comparisons of repeated measurements in linear mixed-effects models followed by Dunn's test with Holm's adjustment, *$P<0.05$ vs. baseline values of MABP or HR, n≥4/group.

**FIGURE 2. Hemodynamic effects of the intravenous administration of apelin-17 (K17F) in conscious hypertensive DOCA-salt rats.** Effects of the i.v. administration of saline in conscious normotensive WKY rats and hypertensive DOCA-salt rats or K17F (20 nmol/kg) in conscious hypertensive DOCA-salt rats on **(A)** MABP and **(B)** HR recorded for 1 hour and respectively expressed in mmHg and bpm. Comparisons of repeated measurements in linear mixed-effects models followed by Dunn's test with Holm's adjustment, *$P<0.05$ vs. baseline values of MABP or HR, n≥5/group.

**FIGURE 3. NOS inhibition and hemodynamic effects of LIT01-196.** **(A)** Effects of the i.v. administration of saline, LIT01-196 (15 nmol/kg) and the simultaneous administration of LIT01-196 and L-NAME (111 µmol/kg) in conscious normotensive SD rats on MABP, recorded for 20 minutes. **(B)** Effects of the i.v. administration of saline, LIT01-196 (3 nmol/kg) and the simultaneous administration of LIT01-196 and L-NAME (111 µmol/kg) in conscious hypertensive DOCA-salt rats on MABP, recorded for 20 minutes. Comparisons of repeated measurements in linear mixed-effects model followed by Dunn's test with Holm's adjustment, *P*<0.05, vs. the baseline value of MABP, n≥5/group.

**FIGURE 4. Hemodynamic effects of the acute subcutaneous administration of 90 nmol/kg LIT01-196 in conscious normotensive WKY and hypertensive DOCA-salt rats.** **(A)** MABP and **(B)** HR recorded for 6 hours after administration of saline or LIT01-196 (90 nmol/kg s.c.) to conscious normotensive WKY rats. **(C)** MABP and **(D)** HR recorded for 7 and 24 hours after administration of saline into conscious normotensive WKY and hypertensive DOCA-salt rats or LIT01-196 (90 nmol/kg s.c.) to conscious DOCA-salt rats. Comparisons of repeated measurements in linear mixed-effects models followed by Dunn's test with Holm's adjustment, *P*<0.05 vs. baseline values of MABP or HR, n≥4/group.

**FIGURE 5. Hemodynamic effects of repeated subcutaneous administration of LIT01-196 in conscious hypertensive DOCA-salt rats.** Effects of repeated daily s.c. administration of saline or LIT01-196 (90 nmol/kg) in conscious hypertensive DOCA-salt rats on **(A)** MABP and **(B)** HR recorded for 6 and 24 hours, and respectively expressed in mmHg and bpm. All measurement started on the day of the fourth s.c. administration. Comparisons of repeated measurements in linear mixed-effects models followed by Dunn's test with Holm's adjustment, *P*<0.05 vs. baseline values of MABP or HR, n≥5/group.

**FIGURE 6. Plasma apelin levels in normotensive WKY and hypertensive DOCA-salt rats**. All measurements were performed in the absence of drug administration. n≥9/group. Comparisons were performed using Student's *t* test. *p<0.05

**FIGURE 7. Metabolic effects of repeated LIT01-196 administrations in DOCA-salt hypertensive rats.** Hypertensive DOCA-salt rats were housed in metabolic cages and received daily injections of saline (grey bar) or LIT01-196 (90 nmol/kg, blue bar) for 4 days. Urine and plasma samples were collected during the 24 hours following the last injection . **(A)** Urine output, **(B)** Urine osmolality, **(C)** Water intake, **(D)** Food intake, **(E)** Plasma sodium, **(F)** Plasma potassium, **(G)** Plasma creatinine, **(H)** Plasma urea nitrogen, **(I)** Plasma glucose, **(J)** Water excretion fraction, **(K)** 24-h sodium excretion, **(L)** 24-h potassium excretion. Comparisons performed using Student's *t* test, *P*<0.05 vs. saline, n=5/group .